\documentclass[runningheads]{llncs}
\usepackage[T1]{fontenc}
\usepackage{graphicx}
\usepackage{amsmath}
\usepackage{multirow}
\newcommand{\repeatthanks}{\textsuperscript{\thefootnote}}

\begin{document}
\title{Supervised Contrastive Learning Framework for Electroencephalography-based Air-writing Recognition}
\author{
Anant Jain\inst{1\thanks{Authors contributed equally to this work.}}\orcidID{0000-0002-7131-8310} \and
Ayush Tripathi\inst{2\repeatthanks}\orcidID{0000-0002-7944-2260}
}
\authorrunning{A. Jain and A. Tripathi}
\institute{
School of Technology, IFIM College, Bangalore, India\\
\email{anant.jain@ifim.edu.in}
\and
Department of Electrical Engineering, Indian Institute of Technology Indore, India\\
\email{ayush.tripathi@iiti.ac.in}
}
\maketitle
\begin{abstract}
Electroencephalography (EEG) - based air-writing recognition offers a human-computer interaction paradigm by decoding neural activity associated with handwriting movements. Despite its potential, reliable EEG-based air-writing recognition remains challenging due to low signal-to-noise ratio and pronounced inter-subject variability. In this study, we examine the use of supervised contrastive learning to improve representation learning for EEG-based air-writing recognition. The analysis is conducted on preprocessed EEG signals and independent component analysis (ICA)--derived neural components obtained from five participants, with trials segmented from $-1$ to $2\,\mathrm{s}$ relative to movement onset. EEGNet and DeepConvNet architectures are evaluated under both conventional cross-entropy training and a supervised contrastive learning framework using a subject-dependent five-fold cross-validation scheme. The results indicate that supervised contrastive learning consistently improves classification accuracy across architectures and feature representations. For preprocessed EEG signals, the mean accuracy increases from 33.45\% to 43.77\% and from 29.14\% to 38.06\% with EEGNet and DeepConvNet, respectively. Using ICA components, higher mean accuracies of 49.21\% and 43.32\% are achieved with EEGNet and DeepConvNet, respectively. These results suggest that the supervised contrastive learning framework offers an efficient extension to existing EEG-based air-writing recognition approaches.

\end{abstract}
\section{Introduction}
\subsection{Background and Related Work}
Air-writing recognition seeks to identify characters written in free space by unconstrained finger or wrist motions \cite{chen2015air-partI,chen2015air-partII}. This offers a natural, touchless input framework for emerging human–computer interaction (HCI) technologies \cite{mitra2007gesture}. In contrast to gesture-based interfaces, air-writing does not require users to adopt artificial gestures, as characters can be formed using conventional writing motions. Air-writing is particularly suitable for wearable computing \cite{patil2016handwriting}, augmented and virtual reality environments \cite{shivener2025writing}, and assistive technologies \cite{vaidya2022air}. Most recent air-writing recognition systems utilize non-neural sensing modalities such as inertial measurement units (IMUs) \cite{tripathi2021sclair,tripathi2022imair}, surface electromyography (EMG) \cite{tripathi2023surfmyoair,tripathi2023tripceair}, or computer vision-based techniques \cite{kim2022writing,wu2024finger}. Despite their robust performance, these modalities depend on peripheral sensors that are sensitive to device-specific characteristics and exhibit limited generalization across users. These limitations motivate the exploration of neural sensing modalities that capture scalp brain signals corresponding to motor execution.

Electroencephalography (EEG) provides a non-invasive method for recording cerebral activity associated with motor planning and execution \cite{falk2023brain}. EEG-based BCI systems are popular due to low cost, portability, and high temporal resolution \cite{jain2022premovnet}. EEG-based brain–computer interface (BCI) systems have been used for motion recognition in various motor-execution tasks, such as biceps curl \cite{saini2023bicurnet}, grasp-and-lift \cite{jain2025esi}, and turn detection \cite{anand2024eeg}. Pei et al. \cite{pei2021online} proposed an EEG-based handwritten letters recognition system using a convolutional neural network (CNN) decoder. Crell et al. \cite{crell2024handwritten} used EEG-based letter kinematics decoding to recognize 10 letters. Various EEG-based features were explored for the recognition of 21 letters by Tripathi et al \cite{tripathi2024neuroair}. However, most EEG-based air-writing recognition methods rely on traditional cross-entropy–based supervised learning. It primarily focuses on optimizing decision boundaries and may struggle to learn robust and discriminative representations. Recently, Supervised contrastive learning (SPL) \cite{khosla2020supervised} has emerged as an efficient alternative, specifically promoting intra-class compactness and inter-class separability within the learned embedding space. The study explores the effectiveness of SPL in air-writing recognition using EEG signals.

\subsection{Objectives and Contributions}
This study aims to integrate supervised contrastive learning and EEG-based air-writing recognition. The objectives of the study are to apply a supervised contrastive learning framework to scalp-recorded EEG signals and evaluate its efficacy in learning discriminative neural representations for air-written character classification.

The key contributions of the study are summarized as follows:
\begin{enumerate}
    \item Developing a supervised contrastive learning framework for EEG-based air-writing recognition, facilitating structured representation learning from scalp EEG signals.

    \item Integration of deep-learning architectures and performance evaluation using processed EEG and ICA-derived feature representations.

    \item Comparative analysis between supervised contrastive learning and conventional cross-entropy–based training in a multi-subject EEG air-writing setting.
    
\end{enumerate}

This study expands contrastive representation learning to EEG-based air-writing recognition and offers a compact yet efficient methodological enhancement for neural HCI systems.

\section{Materials and Methods}

\subsection{Data Description}

In this study, the publicly available \textit{NeuroAiR} dataset \cite{tripathi2024neuroair} is utilized. The dataset comprises EEG recordings from healthy right-handed participants performing an air-writing task. Data corresponding to five subjects are analyzed in this study. During the recording sessions, participants were seated comfortably and instructed to remain relaxed, with their right elbow resting on a table to minimize upper-arm movement and reduce motion-related artifacts. A custom graphical user interface developed with the Tkinter module in Python was used to present visual cues for uppercase English letters. Participants interacted with the interface using their non-dominant hand, while writing the prompted characters in free space using their index finger.

EEG signals were recorded using a 31-channel LiveAmp EEG system (Brain Products GmbH, Germany), with electrodes placed according to the international 10--20 system using an EasyCap electrode cap. The signals were recorded with a sampling rate of 500 Hz. Electrode impedance was maintained below 20 k$\Omega$ throughout the recording sessions to ensure adequate signal quality. At the start of each trial, the participant initiated the task by pressing the space bar, after which a single character was presented on the screen. The participant was instructed to write the displayed character in free space using the index finger. The trial was ended with the press of the space bar again. Each participant completed 100 repetitions of the full set of 26 uppercase English letters, which yielded a total of 2600 samples per subject. Each letter was recorded individually, and short breaks were provided after every five sets of recordings to reduce fatigue.

\subsection{EEG Preprocessing}
EEG preprocessing was performed using EEGLAB \cite{delorme2004eeglab} toolbox in MATLAB software. The preprocessing enhanced the EEG signal quality by reducing artifacts and extracting neural feature representations. Raw EEG signals were common-average-referenced. Subsequently, the signals were bandpass-filtered between 0.5 and 45 Hz using zero-phase, non-causal finite-impulse-response (FIR) filters to remove slow drifts and high-frequency noise. Further, independent component analysis (ICA) was applied to decompose the EEG signals into statistically independent source components. Under the ICA model, the multichannel EEG signal is represented as a linear combination of independent components, enabling the separation of neural sources from non-neural artifacts such as eye blinks and muscle activity. Artifact-related components were identified using the ICLabel algorithm implemented in EEGLAB. The components corresponding to the ocular or muscular artifacts, with a confidence score exceeding 0.8, were removed. Clean EEG signals were then reconstructed by recombining the remaining components using a modified mixing matrix.

In this study, only two feature representations were retained for subsequent analysis: \textit{Preprocessed EEG signals}, corresponding to artifact-free EEG data, and \textit{ICA component time series}, representing spatially filtered neural source activities. Both representations were segmented into fixed-length trials corresponding to the air-writing task and normalized using z-score normalization prior to model training.

\subsection{Supervised Contrastive Learning Framework}
To enhance the discriminative power of EEG-based air-writing representations, this study employs a \textit{supervised contrastive learning (SCL)} framework.

Let a mini-batch of $N$ labeled EEG trials be denoted as
$\{(x_i, y_i)\}_{i=1}^{N}$, where $x_i$ represents an EEG trial corresponding to an air-written character and $y_i \in \{1, \dots, C\}$ indicates its class label. Each input trial is mapped into a latent representation via an encoder network,
\begin{equation}
r_i = \text{Enc}(x_i),
\end{equation}
where $\text{Enc}(\cdot)$ corresponds to a deep neural network encoder. The latent representation $r_i$ is then normalized and passed through a projection head to obtain
\begin{equation}
z_i = \text{Proj}(r_i),
\end{equation}
where $\text{Proj}(\cdot)$ denotes a nonlinear mapping implemented using a fully connected layer.

The supervised contrastive loss promotes samples belonging to the same class to form compact clusters in the embedding space while enforcing separation between samples from different classes. For a given anchor sample $i$, let $P(i)$ denote the set of indices corresponding to samples in the mini-batch that share the same class label as $i$, and let $A(i)$ denote the set of all samples in the batch except $i$. The supervised contrastive loss is defined as
\begin{equation}
\mathcal{L}_{\text{sup}} =
\sum_{i=1}^{N}
\frac{-1}{|P(i)|}
\sum_{p \in P(i)}
\log
\frac{\exp(z_i \cdot z_p / \tau)}
{\sum_{a \in A(i)} \exp(z_i \cdot z_a / \tau)},
\end{equation}
where $\tau$ is a temperature scaling parameter and $(\cdot)$ denotes the inner product between normalized embedding vectors.

The minimization of the loss function explicitly promotes \textit{intra-class similarity} and \textit{inter-class dissimilarity} within the learned embedding space. In contrast to the conventional cross-entropy loss, which focuses on optimizing decision boundaries, the supervised contrastive loss directly shapes the geometry of the latent space. Hence, it improves representation robustness in the presence of inter-trial and inter-subject variability commonly observed in EEG signals.

The overall learning strategy follows a \textit{two-stage training protocol}. In the first stage, the encoder and projection head are jointly optimized using the supervised contrastive loss to learn discriminative EEG representations. In the second stage, the projection head is discarded, and a linear classification head is trained on the frozen encoder outputs using cross-entropy loss. This strategy ensures that, at inference time, the model complexity remains identical to that of a standard classification network while leveraging contrastively learned representations.

\subsection{Model Architectures}

In the present study, two established convolutional neural network architectures— \textit{EEGNet} \cite{lawhern2018eegnet} and \textit{DeepConvNet} \cite{schirrmeister2017deep} —are utilized for EEG-based airwriting recognition. These models were selected due to their demonstrated effectiveness in EEG decoding tasks. Both architectures are evaluated \textit{with and without the supervised contrastive learning (SCL) framework} across all five participants.

\subsubsection{EEGNet}

EEGNet is a compact convolutional neural network specifically designed for EEG-based brain–computer interface applications. The architecture seeks to extract interpretable spatial–temporal features from multichannel EEG signals with constrained training data. EEGNet consists of two primary stages. In the first stage, temporal convolutions are applied to the input EEG signal using two-dimensional convolutional filters to learn frequency-selective representations. This is followed by a depthwise convolution operation that acts as a spatial filter across EEG channels, enabling the model to learn channel-wise spatial patterns associated with neural activity. Batch normalization and the exponential linear unit (ELU) activation function are applied to stabilize training and introduce nonlinearity, respectively. Average pooling and dropout layers are utilized to reduce dimensionality and avoid overfitting. In the second stage, separable convolutions are employed to enhance the representation learning while maintaining a minimal parameter count. The output feature maps are flattened and passed to a fully connected classification layer with softmax activation in the cross-entropy–based training setup. In the supervised contrastive learning framework, the softmax classifier is replaced by a projection head during the representation learning stage, while the encoder remains unchanged.

\begin{table}[!t]
\centering
\caption{EEGNet Model Architecture}
\label{tab:eegnet_arch}
\begin{tabular}{ll}
\hline
\textbf{Layer} & \textbf{Configuration} \\
\hline
Input & EEG signal ($31 \times 1500$) \\
Conv2D & 8 filters, kernel $(1 \times 64)$ \\
Batch Normalization & -- \\
Depthwise Conv2D & kernel $(31 \times 1)$, depth multiplier = 2 \\
Batch Normalization & -- \\
Activation & ELU \\
Average Pooling & $(1 \times 4)$ \\
Dropout & 0.5 \\
Separable Conv2D & kernel $(1 \times 16)$ \\
Batch Normalization & -- \\
Activation & ELU \\
Average Pooling & $(1 \times 8)$ \\
Flatten & -- \\
Dense (CE setup) & 26 units, softmax \\
Dense (SCL setup) & Projection head (during Stage 1) \\
\hline
\end{tabular}
\end{table}

\subsubsection{DeepConvNet}

DeepConvNet is a convolutional neural network architecture comprising a sequence of convolutional and pooling blocks that hierarchically extract temporal and spatial features from EEG signals. The first block of DeepConvNet consists of consecutive convolutional layers that perform temporal filtering, followed by spatial filtering across EEG channels. Subsequent blocks consist of convolutional layers with increasing numbers of filters, each followed by batch normalization, ELU activation, and max-pooling operations. This hierarchical structure enables the model to learn progressively abstract representations of EEG activity related to air-writing movements. In the conventional classification setting, the output of the final convolutional block is flattened and fed into a fully connected softmax layer trained using cross-entropy loss. In the supervised contrastive learning configuration, the DeepConvNet encoder is coupled with a projection head during the first training stage, as in the EEGNet-based SCL setup.

\begin{table}[!t]
\centering
\caption{DeepConvNet Model Architecture}
\label{tab:deepconvnet_arch}
\begin{tabular}{ll}
\hline
\textbf{Layer} & \textbf{Configuration} \\
\hline
Input & EEG signal ($31 \times 1500$) \\
Conv2D & 25 filters, kernel $(1 \times 5)$ \\
Conv2D & 25 filters, kernel $(31 \times 1)$ \\
Batch Normalization & -- \\
Activation & ELU \\
Max Pooling & $(1 \times 2)$ \\
Conv2D & 50 filters, kernel $(1 \times 5)$ \\
Batch Normalization & -- \\
Activation & ELU \\
Max Pooling & $(1 \times 2)$ \\
Conv2D & 100 filters, kernel $(1 \times 5)$ \\
Batch Normalization & -- \\
Activation & ELU \\
Max Pooling & $(1 \times 2)$ \\
Conv2D & 200 filters, kernel $(1 \times 5)$ \\
Batch Normalization & -- \\
Activation & ELU \\
Max Pooling & $(1 \times 2)$ \\
Flatten & -- \\
Dense (CE setup) & 26 units, softmax \\
Dense (SCL setup) & Projection head (during Stage 1) \\
\hline
\end{tabular}
\end{table}

For both EEGNet and DeepConvNet, the supervised contrastive learning framework is implemented by treating the convolutional backbone as an \textit{encoder network}. During the first stage of training, the encoder is enhanced with a projection head and optimized using supervised contrastive loss to learn discriminative latent representations. The projection head comprises of a single fully connected layer with 128 neurons and ReLU activation. The encoder and projection head parameters are jointly optimized by minimizing the supervised contrastive loss using the Adam optimizer. During the classification stage, the projection head is discarded and substituted by a fully connected output layer consisting 26 neurons with softmax activation. A dropout rate of 0.5 is applied to the classifier to reduce overfitting, and the classification network is trained using the standard cross-entropy loss.

To evaluate the impact of contrastive representation learning, each model is trained and evaluated in two configurations:
\begin{enumerate}
    \item Standard supervised learning, using cross-entropy loss alone, and
    \item Supervised contrastive learning–based training, following the two-stage protocol.
\end{enumerate}

All experiments are uniformly conducted across the five participants, facilitating a direct comparison of architectures and training strategies while isolating the effect of supervised contrastive learning on EEG-based air-writing recognition.

\section{Results and Discussion}
\subsection{Experimental Details}

In this study, experiments were conducted using \textit{preprocessed EEG signals} and \textit{independent component analysis (ICA) component time series} as input features for EEG-based air-writing recognition. For each trial, EEG segments were extracted from a temporal window spanning \(-1\,\mathrm{s}\) to \(2\,\mathrm{s}\) relative to the onset of the air-writing task. This window was selected to capture both pre-movement neural activity and execution-related dynamics. Since EEG signals were recorded at a sampling rate of \(500\,\mathrm{Hz}\), each trial segment consisted of \(1500\) time samples. For shorter-duration trials, zero-padding was applied to ensure a fixed-length representation across all samples. The resulting input feature matrices were of dimension \(\kappa \times 1500\), where \(\kappa = 31\) for both preprocessed EEG signals and ICA component time series. All input data were normalized using z-score normalization to achieve a mean of 0 and a variance of 1.

A \textit{user-dependent 5-fold cross-validation} technique was employed to evaluate the model's performance. For each participant, the dataset was split into five mutually exclusive folds. Four folds are used for training, while one fold is reserved for testing in each iteration. This procedure was repeated until each fold functioned as the test set once. Within each training fold, a subset of the data was allocated as a validation set for early stopping. The performance metrics were averaged across the folds to obtain subject-specific results. All deep learning models were trained with a \textit{mini-batch size of 32}. Model optimization was performed using the Adam optimizer with early stopping configured to a patience of 20 on validation accuracy. The final reported results correspond to the average performance across all five participants, ensuring a uniform evaluation of the proposed and baseline models.

\subsection{Results}

\begin{table}[t]
\centering
\caption{Mean accuracy score across the participants using pre-processed EEG and ICA components time series data in cross-entropy loss and supervised contrastive loss settings.}
\scalebox{1}{
\centering
\begin{tabular}{ccccc}
\hline \hline
\multicolumn{5}{c}{\textbf{Pre-processed   EEG}}                                                                                                \\ \hline\hline
\multirow{2}{*}{\textbf{participants}} & \multicolumn{2}{c}{\textbf{Cross-Entropy Loss}} & \multicolumn{2}{c}{\textbf{Supervised Contrastive Loss}} \\ \cline{2-5} 
                                   & \textbf{EEGNet}      & \textbf{DeepConvNet}     & \textbf{EEGNet}          & \textbf{DeepConvNet}          \\ \hline
Sub01                              & 33.58                & 33.85                    & 43.88                    & 39.58                         \\
Sub02                              & 30.62                & 24.00                    & 38.73                    & 34.54                         \\
Sub03                              & 35.31                & 28.23                    & 55.88                    & 44.31                         \\
Sub04                              & 42.69                & 36.85                    & 50.92                    & 43.65                         \\
Sub05                              & 25.04                & 22.77                    & 29.42                    & 28.23                         \\ \hline
\textbf{Average}                   & 33.45                & 29.14                    & \textbf{43.77}           & 38.06                         \\ \hline \hline
\multicolumn{5}{c}{\textbf{ICA Components}}                                                                                                     \\ \hline \hline
\multirow{2}{*}{\textbf{participants}} & \multicolumn{2}{c}{\textbf{Cross-Entropy Loss}} & \multicolumn{2}{c}{\textbf{Supervised Contrastive Loss}} \\ \cline{2-5} 
                                   & \textbf{EEGNet}      & \textbf{DeepConvNet}     & \textbf{EEGNet}          & \textbf{DeepConvNet}          \\ \hline
Sub01                              & 35.04                & 35.62                    & 49.65                    & 44.77                         \\
Sub02                              & 35.04                & 29.31                    & 44.46                    & 38.88                         \\
Sub03                              & 41.12                & 30.62                    & 59.04                    & 50.35                         \\
Sub04                              & 43.62                & 42.12                    & 57.88                    & 49.85                         \\
Sub05                              & 27.65                & 24.73                    & 35.00                    & 32.73                         \\ \hline
\textbf{Average}                   & 36.49                & 32.48                    & \textbf{49.21}           & 43.32                         \\ \hline
\end{tabular}
}
\end{table}
Table~3 summarizes the \textit{mean classification accuracy} obtained across five participants using \textit{preprocessed EEG signals} and \textit{ICA component time series}, evaluated under \textit{cross-entropy (CE)} and \textit{supervised contrastive learning (SCL)} training paradigms with EEGNet and DeepConvNet architectures. For preprocessed EEG inputs, models trained using supervised contrastive learning consistently outperform their cross-entropy-trained counterparts across all participants and architectures. Averaged across participants, EEGNet trained with CE achieves a mean accuracy of 33.45\%, which increases to 43.77\% when trained using the SCL framework. A similar trend is observed for DeepConvNet, with mean accuracy increasing from 29.14\% (CE) to 38.06\% (SCL). Mean accuracy analysis shows that the SCL framework yields significant improvements in classification performance. For Sub03, an increase in accuracy from 35.31\% (CE) to 55.88\% (SCL) is observed with the EEGNet decoder. This shows the efficiency of supervised contrastive representation learning for class separability. Overall, EEGNet consistently outperforms DeepConvNet on preprocessed EEG data across both training strategies.

Improvement in classification performance is observed when ICA component time series were utilized as input features across training paradigms and architectures. With cross-entropy training, the mean accuracy increases to 36.49\% for EEGNet and 32.48\% for DeepConvNet. Further, the supervised contrastive learning framework improved the accuracy to 49.21\% and 43.32\% for EEGNet and DeepConvNet, respectively. For ICA-based features, the improvements in performance using the SCL framework are significantly greater than those with preprocessed EEG signals. The highest average accuracy of 49.21\% is achieved with EEGNet and the SCL framework using ICA-features, an improvement of $\approx$12.7\% over the cross-entropy baseline. SCL framework consistently outperforms the traditional cross-entropy-based training framework. Furthermore, ICA-derived features outperform preprocessed EEG signals, which shows the significance of spatially decomposed neural representations for EEG-based air-writing recognition. Mean accuracy scores demonstrate that integrating supervised contrastive learning with deep learning classifiers yields better discriminative latent representations for EEG-based air-writing recognition.

\section{Discussion}

The experimental results demonstrate the efficiency of the supervised contrastive learning framework for EEG-based air-writing recognition in comparison to cross-entropy-based training. The improvements are observed in both preprocessed EEG signals and ICA component-based features, using both the EEGNet and DeepConvNet classifiers. ICA-based features achieve higher classification accuracy than preprocessed EEG signals. This indicates that independent component representations provide better discriminative input for learning neural embeddings. Across both training configurations, EEGNet outperforms DeepConvNet, highlighting the importance of compact, EEG-specific architectural inductive biases. Overall, the classification analysis shows the robustness and efficiency of the supervised contrastive learning framework for EEG-based air-writing recognition.

\section{Conclusion}

This study introduced the supervised contrastive learning framework for EEG-based air-writing recognition. The preprocessed EEG signals and ICA-derived neural components were used as input features. The proposed framework integrates contrastive representation learning with EEGNet and DeepConvNet classifiers to improve classification performance. Experimental results demonstrate the superiority of supervised contrastive learning over conventional cross-entropy-based training. A significant performance improvement while using ICA component features. EEGNet outperforms DeepConvNet, showing the effectiveness of compact EEG-specific architectures. Overall, the framework provides an effective extension to existing EEG-based air-writing recognition frameworks.

\bibliographystyle{splncs04}
\bibliography{MBCC2026}

\end{document}